\documentclass[6ptpt, preprint2]{aastex}
\usepackage[]{natbib}
\begin{document}

\title{Comments on the Redshift Distribution of 44,200 SDSS Quasars: Evidence for Predicted Preferred Redshifts?}

\author{M.B. Bell\altaffilmark{1}}
\altaffiltext{1}{Herzberg Institute of Astrophysics,
National Research Council of Canada, 100 Sussex Drive, Ottawa,
ON, Canada K1A 0R6;
morley.bell@nrc.ca}

\begin{abstract}

A Sloan Digital Sky Survey (SDSS) source sample containing 44,200 quasar redshifts is examined. Although arguments have been put forth to explain some of the structure observed in the redshift distribution, it is argued here that this structure may just as easily be explained by the presence of previously predicted preferred redshifts.

\end{abstract}

\keywords{galaxies: active - galaxies: distances and redshifts - galaxies: quasars: general}

\section{Introduction.}

It was realized early in the SDSS survey \citep{ric02} that quasar redshifts near z = 2.7 would be under-sampled simply because at that redshift the colors of stars resemble those of quasars and it would be difficult to keep the quasar identification efficiency high. Based on the knowledge gained during the commissioning period, and the results from the first 2555 quasars, the target selection algorithm was refined to the point where it was felt by 2000 November that it was appropriate for use in obtaining a statistical sample of quasars \citep{ric02}. However, after additional observations were carried out, and many more sources with redshifts above z = 3 were obtained, it became apparent that there was a second dip in the redshift distribution at z = 3.5. This was again deemed to be a problem with the source selection algorithm, although it is not clear how this was determined, since a real source deficiency at this redshift would presumably give the same result. It may, however, have been decided from the evidence for a trough visible in the $i$-redshift plot \citep[Fig 3]{sch03} which, when redshifts are assumed to be purely cosmological, can only be explained by a selection effect. On the other hand, it is exactly what would be expected if the redshifts are intrinsic. However, further changes to the target selection algorithm were made and it was \em finalized \em a second time on 2001 August 24. Spectra taken since that date were deemed to constitute the basis for the statistical sample of SDSS quasars \citep{ric02}. 
The first release of SDSS quasar redshift data (DR1) contained the 16,713 redshifts measured prior to 2001 October \citep{sch03}. Although the target selection algorithm was being refined throughout most of the period the DR1 observations were made, the sample considered here contains an additional $\sim 27,500$ redshifts that were obtained since 2001 October, after the algorithm was finalized.

The primary SDSS quasar science goals were to study (1), the evolution of the quasar luminosity function and (2), the spatial clustering of quasars, as a function of redshift. However, a reasonably complete statistical sample of quasar redshifts can also be used to examine the long-standing controversy surrounding the very nature of quasar redshifts. Peaks in the distribution of quasar redshifts have long been claimed as possible evidence for the existence of preferred redshifts \citep{kar71,kar77,bur01,bel02a,bel02c}. However, only source samples with a few hundred redshifts have been available to test this claim. Now that surveys containing tens of thousands of quasar redshifts are available, if selection effects can be properly accounted for, although this will not be easy, any peaks still visible in the redshift distribution could be a much more reliable indicator for the existence of preferred redshifts. Alternatively, if predicted redshift peaks do not appear, it can mean either that intrinsic redshifts do not exist, or that they have been smeared out by a significant Doppler component.

 There are currently only two models that have been proposed that clearly define the preferred redshift values expected. It is of interest now to see if one of these models might equally well explain the observed redshift distribution of 44,200 SDSS quasars.

\section{Preferred Redshift Models}

Over 30 years ago it was pointed out \citep{kar71,kar77} that the relation

 $\Delta$log(1+z) = 0.089 ------------------------- (1)

appears to predict redshifts that are preferred in the emission line redshift distribution of the 574 strong quasars detected in early surveys. Most of these redshifts were between z = 0 and z = 2. Since that time much work has been carried out in an attempt to find further evidence to support this claim \citep{bur01}. A significant amount of evidence has also been accumulated supporting the argument that high-redshift QSOs are ejected from nearby active galaxies, and that the redshifts of these objects are largely intrinsic \citep{arp98,arp99,bur95,bur97a,bur97b,chu98,pie94,lop02}. Below z = 4.8, eqn 1 predicts preferred redshifts of z = 0.061, 0.3, 0.6, 0.96, 1.41, 1.96, 2.63, 3.45, and 4.47 \citep{bur01}.

In addition to the model defined by eqn 1, a second intrinsic redshift model has been reported \citep{bel02a,bel02c} that predicts intrinsic redshifts periodic in z. This model resulted primarily from the study of the 15 high-redshift QSOs located around the Seyfert galaxy NGC 1068 \citep{bel02a,bel02b,bel02c,bel03a} which also assumed that the QSOs had been ejected from the Seyfert galaxy.
In this model the high redshifts of the QSOs must contain large intrinsic components if they are attached to NGC 1068 which has a cosmological redshift of z = 0.0038. 

 From the combined results of the study of QSOs near NGC 1068, and the results obtained in an earlier one \citep{bur90}, it was found that the intrinsic redshift components in quasars (z$_{iQ}$) can be defined by the relation

z$_{iQ}$ = 0.62[$N$ - $M_{N}$] ---------------- (2)

where $N$ is an integer, and $M_{N}$ is a function of a second quantum number $n$, \citep{bel02c,bel03a}. The intrinsic redshifts predicted by eqn 2 are listed in Table 1, for values less than z = 4.5.


\begin{deluxetable}{cccccccc}
\tabletypesize{\scriptsize}
\tablecaption{Intrinsic Redshifts Predicted by Eqn 2 for z$_{iQ} < 4.5$. \label{tbl-1}}
\tablewidth{0pt}
\tablehead{
\colhead{($n$)} & \colhead{z$_{iQ}[N=1,n$]} & \colhead{z$_{iQ}[N=2,n$]} & \colhead{z$_{iQ}[N=3,n$]} &  \colhead{z$_{iQ}[N=4,n$]}
 & \colhead{z$_{iQ}[N=5,n$]} & \colhead{z$_{iQ}[N=6,n$]} & \colhead{z$_{iQ}[N=7,n$]}
}
\startdata

0 & 0.620 & 1.240 & 1.860 & 2.48 & 3.10 & 3.72 & 4.340 \\
1 & 0.558 & 1.178 & 1.798 & 2.418 & 3.038 & 3.658 & 4.278 \\
2 & 0.496 & 1.054 & 1.488 & 1.178 \\
3 & 0.434 & 0.868 & 0.558 \\
4 & 0.372 & 0.620 \\
5 & 0.310 & 0.310 \\
6 & 0.248 \\
7 & 0.186 \\
8 & 0.124 \\
9 & 0.062 \\

\enddata 
\end{deluxetable}



\begin{figure}
\hspace{-1.0cm}
\vspace{-2.0cm}
\epsscale{1.0}
\plotone{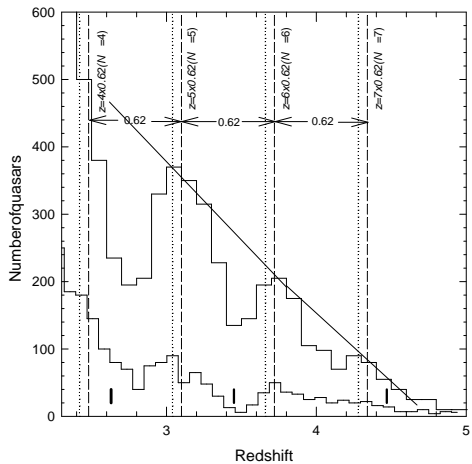}
\caption{\scriptsize{(upper histogram)Distribution of approximately 5000 SDSS quasars between z = 2.3 and z = 5. (lower histogram) Distribution of quasar redshifts in the DR1 catalog. Solid bars near the bottom indicate redshifts predicted by eqn 1. Vertical dashed and dotted lines indicate redshifts predicted by eqn 2. \label{fig1}}}
\end{figure}


\section{Results for the SDSS High-Redshift Data}

In models involving the ejection of QSOs from galaxies, in addition to the intrinsic component, Doppler components due to randomly directed ejection velocities and the Hubble flow will be present in the measured redshifts. An estimate of the size of the ejection velocities expected in such cases was obtained in the study of the 15 QSOs near NGC 1068 \citep{bel02a,bel02c} where ejection velocities up to $\sim 25,000$ km s$^{-1}$ were found. When a cosmological component is added to this, even peaks separated by redshifts of z = 0.62 can easily be smeared out \citep{bel04}.

Equation 1 predicts that the spacing of the intrinsic redshifts increases as the intrinsic redshifts increase. Thus the separation between the predicted lines will be greater at high redshifts and are more likely to be free of the confusion that can occur at lower redshifts. Also, this redshift range above z = 2 has not previously contained a sufficiently high number of redshifts to allow a meaningful study. It now offers the opportunity to examine a completely different redshift range as well as a vastly more complete and independent sample, keeping in mind that there still may be selection effects in this portion of the data. It is important to realize as well, however, that \em selection effects that miss sources can only affect those regions of the distribution where sources actually exist. \em Therefore it will not be easy to sort out exactly how much of a low-density valley in the redshift distribution is real and how much is due to the selection effect.

For the quasars listed in the DR1 catalog, the redshift distribution between z = 2.3 and z = 5 is plotted in the lower histogram in Fig. 1. There is clearly a drop in the number of quasars at z = 2.75.

In this same redshift range the distribution of approximately 5000 quasar redshifts in the sample containing 44,200 redshifts considered here \citep{ogu03} is given by the upper histogram in Fig 1. Most of the redshifts in this distribution have been obtained after several iterations of the target selection algorithm, and this process has clearly removed the narrow dip at z = 2.75. However, prominent peaks and valleys still remain. In fact, in some cases, the peak-to-valley ratio may be even higher than in the DR1 data.

The locations of the preferred values predicted by eqn 1 are indicated in Fig 1 by short, bold vertical bars near the lower axis. There is clearly no agreement between these redshifts and the peaks in the distribution. This model is therefore not supported by these high-redshift data. 

The vertical dashed and dotted lines in Fig. 1 indicate the locations of the intrinsic components predicted in this redshift range by eqn 2. Not only is there excellent agreement between these values and the peaks in the redshift distribution, there are also three clear valleys between the peaks. Valleys are expected here because no redshifts are predicted. They are not expected to drop to zero, however, because the preferred values will be smeared out by Doppler components. Not only are the peaks periodic in z = 0.62, they are pure harmonics above z = 0. While possible causes for the valleys at z = 2.7 and z = 3.5 have been suggested, no reason to suspect a valley at z = 4.03 has been suggested. It is concluded here that eqn 2 appears to explain the redshift distribution in Fig 1 better than the other arguments that have so far been proffered. Furthermore, although it is possible to suggest explanations for the valleys at z = 2.7 and 3.5, \em there is no way to prove that the reasons claimed are their true cause. If there are actually no redshifts in the valleys, the selection effects raised above cannot affect the result. \em

In Fig 1 it needs to be kept in mind from the magnitudes involved, that if these high redshift objects have been ejected from nearby low-redshift galaxies, and are therefore still associated with them, they must be intrinsically several magnitudes fainter than normal galaxies. Because they are therefore too faint to be detected at large distances, their cosmological, or distance redshift components will be quite small relative to their intrinsic redshifts. In this model the distribution in Fig 1 is therefore essentially an intrinsic redshift one. If, as is assumed in this model, quasars are ejected from active galaxies, and these galaxies are distributed uniformly in space, the same should be true for quasars. Distant ones will not be detectable, however, with current sensitivities. On the other hand, those with lower intrinsic redshifts, which have been found to be more luminous than those with high intrinsic redshifts \citep{bel02a,bel02c}, will be detectable to greater cosmological distances.


\begin{figure}
\hspace{-1.0cm}
\vspace{-2.0cm}
\epsscale{1.0}
\plotone{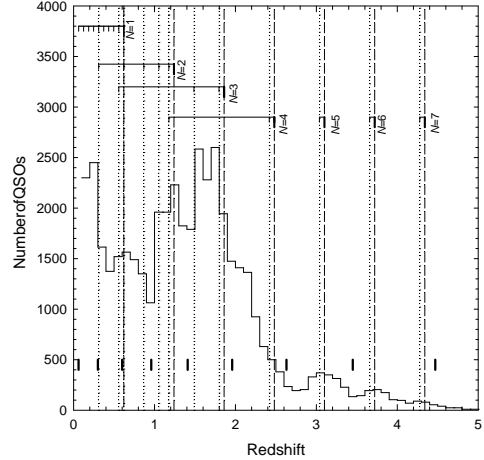}
\caption{\scriptsize{Distribution of 44,200 SDSS quasar redshifts. Solid bars near the bottom indicate redshifts predicted by eqn 1. Vertical dashed and dotted lines indicate redshifts predicted by eqn 2. \label{fig2}}}
\end{figure}



\begin{figure}
\hspace{-1.0cm}
\vspace{-2.0cm}
\epsscale{1.0}
\plotone{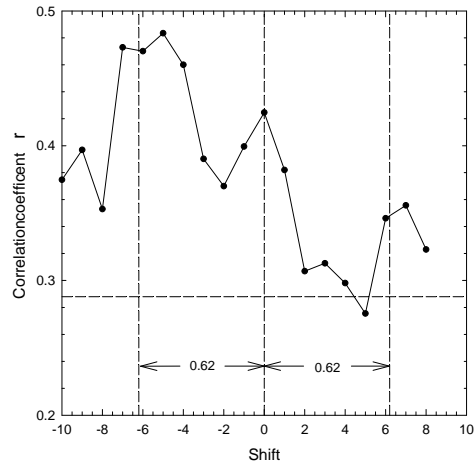}
\caption{\scriptsize{Correlation coefficient r between the distribution of 44,200 SDSS quasar redshifts and the distribution of lines predicted using eqn 2. The horizontal dashed line indicates the level above which r is considered significant. \label{fig3}}}
\end{figure}


\section{The Full Sample of 44,200 Redshifts}

The entire sample containing 44,200 redshifts is plotted in Fig 2. As in Fig 1, from eqn 2, the vertical dashed lines indicate the highest redshift level in each $N$-group and the dotted lines indicate the lower levels inside each group. Where the region above z = 2.3 contains only two lines in each group, as can be seen in Fig. 2 and in Table 1, below z = 2 the density of predicted intrinsic redshift lines inside each group increases significantly.

Immediately evident in Fig 2 is the fact that the density of SDSS redshifts in the lower half of the sample is much higher than in the upper half, although part of this difference can be attributed to the low-redshift sample with $i$-mag cut-off at $i$ = 19.1. Similarly, there is a much higher density of predicted lines in the lower region, especially for the model described by eqn 2. Although, for eqn 2, the higher line density below z = 2 introduces some confusion, there is still evidence for peaks that correspond to the first three $N$-groups located at, or near, z = 0.62, 1.24, and 1.86, depending on the line distribution inside each group. Only at z = 2.48, where the source density is rising steeply, is there no obvious peak, although the DR1 distribution in Fig 1 does show one. The lack of a clear peak at z = 2.48 is not surprising since the survey portion with $i$-mag cut-off at $i$ = 19.1 found mainly sources below z $\sim$2.2. This has contributed to the sharp drop in source density above z = 2.2, and such an abrupt change in level can easily mask a weak peak. 

In Fig 3 the correlation coefficient r, between the actual redshift distribution and the distribution of redshifts predicted using eqn 2 is plotted versus shift. It shows a correlation coefficient of 0.42 for zero shift. For a distribution containing 45 samples as here, r greater than 0.288 is considered significant. However, more importantly, as the two distributions are shifted relative to each other, r first decreases and then increases again at $\pm6$ bin shifts (one bin = 0.1 in redshift) corresponding to z = $\pm0.6$, clearly indicating the presence of peaks in the distribution that are periodic in z = 0.62. That the highest value of r is obtained for a shift of -0.6 is not surprising since it corresponds to the situation where the highest intrinsic line density (z $\sim 1.2$) is aligned with the highest peak in the quasar redshift distribution (z $\sim 1.8$). This further confirms that the periodic peaks in the distribution below z = 2 are also contributing to the correlation coefficient. The fact that r remains high over many shifts is due to the broad density correlation between the upper and lower halves of the data.

 There was no significant correlation found between the SDSS redshift distribution and the distribution of intrinsic redshifts predicted by eqn 1. This is not surprising since, in contrast to what was found for eqn 2, the intrinsic lines predicted by eqn 1 coincide with only one peak in the SDSS distribution (z = 0.6), and there is little correlation between the source and line densities in the upper and lower halves of the distributions. It therefore must be concluded that the model described by eqn 1 is not confirmed by the SDSS redshift data. Furthermore, while it has previously been demonstrated that the intrinsic redshifts predicted by eqn 2 also show a significant correlation with both the redshift distribution of the 574 strong quasars found in early surveys \citep{bel02c,bel03b}, and the redshift distribution of the QSOs located around NGC 6212 \citep{bel03b}, there was also no significant correlation found in these cases using eqn 1. 

\section{Conclusions}

In this examination of 44,200 SDSS quasar redshifts it has been found that there are at least 5 peaks in the redshift distribution that correspond to the pure harmonics of z = 0.62 predicted by eqn 2. Although valid arguments have been put forward to explain the redshift valleys at z = 2.7 and 3.5, there is no proof yet that they have correctly explained the causes of these low source counts, since eqn 2 predicts low densities at both these redshifts, reducing the effect of any selection effect. Eqn 2 also predicts the low source count seen near z = 4.03 where no selection effect has yet been proposed.

It is concluded here that the intrinsic redshift model given by eqn 2 can better explain the structure present in the distribution of these 44,200 SDSS quasar redshifts than can the selection effects suggested by others. SDSS investigators are urged to keep this in mind if they hope to insure that real density fluctuations are not inadvertently removed. If we are to obtain a truly reliable statistical sample, adjustments that affect the SDSS redshift distribution must not involve the assumption that quasar redshifts are purely cosmological. 

\section{Acknowledgements} 

Funding for the creation and distribution of the SDSS Archive has been provided by the Alfred P. Sloan Foundation, the Participating Institutions, the National Aeronautics and Space Administration, the National Science Foundation, the US Department of Energy, the Japanese Monbukagakusho, and the Max Planck Society. The SDSS website is http://www.sdss.org/. The SDSS is managed by the Astrophysical Research Consortium (ARC) for the Participating Institutions. The Participating Institutions are The University of Chicago, Fermilab, the Institute for Advanced Study, The Japan Participation Group, The Johns Hopkins University, Los Alamos National Laboratory, the Max-Planck-Institute for Astronomy (MPIA), the Max-Planck-Institute for Astrophysics (MPA), New Mexico State University, University of Pittsburgh, Princeton University, the United States Naval Observatory, and the University of Washington.


\clearpage

\begin{thebibliography}



\bibitem[Arp(1998)]{arp98} Arp, H. 1998, \apj, 496, 661
\bibitem[Arp(1999)]{arp99} Arp, H. \apj 1999, 525, 594
\bibitem[Bell(2002a)]{bel02a} Bell, M.B. 2002a, \apj, 566, 705
\bibitem[Bell(2002b)]{bel02b} Bell, M.B. 2002b, \apj, 567, 801
\bibitem[Bell(2002c)]{bel02c} Bell, M.B. 2002c, Preprint at http://xxx.lanl.gov/astro-ph/0208320
\bibitem[Bell and Comeau(2003a)]{bel03a} Bell, M.B., and Comeau, S.P. 2003a, Preprint at http://xxx.lanl.gov/astro-ph/0305060
\bibitem[Bell and Comeau(2003b)]{bel03b} Bell, M.B., and Comeau, S.P.2003b, Preprint at http://xxx.lanl.gov/astro-ph/0306042
\bibitem[Bell(2004)]{bel04} Bell, M.B. 2004, (in preparation)

\bibitem[Burbidge(1995)]{bur95} Burbidge, E.M. 1995, \aap, 298, L1
\bibitem[Burbidge(1997a)]{bur97a} Burbidge, E.M., 1997a, \apj, 484, L99
\bibitem[Burbidge(1997b)]{bur97b} Burbidge, E.M., and Burbidge, G. 1997b, \apj, 477, L13
\bibitem[Burbidge and Hewitt(1990)]{bur90} Burbidge, G. and Hewitt, A. 1990, \apj, 359, L33
\bibitem[Burbidge and Napier (2001)]{bur01} Burbidge, G. and Napier, W.M. 2001, \aj, 121, 21
\bibitem[Chu et al.(1998)]{chu98} Chu, Y. et al. 1998, \apj, 500, 596


\bibitem[Karlsson(1971)]{kar71} Karlsson, K.G. 1971, \aap, 13, 333
\bibitem[Karlsson(1977)]{kar77} Karlsson, K.G. 1977, \aap, 58, 237

\bibitem[Lopez-Corredoira and Gutierez(2002)]{lop02} Lopez-Corredoira, M. \& Gutierez, C.M. 2002, \aap, 390, L15
\bibitem[Oguri et al. (2003)]{ogu03} Oguri, M. et al. 2003, ApJ, (in press), preprint at http://xxx.lanl.gov/astro-ph/0312429
\bibitem[Pietsch et al.(1994)]{pie94} Pietsch, W., Volger, A., Kahabka, P.,Jain, A., and Klein, V. 1994, \aap, 284, 386
\bibitem[Richards et al.(2002)]{ric02} Richards, G. T. et al. 2002, \aj, 123, 2945
\bibitem[Schneider et al.(2003)]{sch03} Schneider et al. 2003, \aj, 126, 2579
\bibitem[Stoughton et al.(2002)]{sto02} Stoughton, C. et al. 2002, \aj, 123, 485

\end{thebibliography}
\end{document}